# Using Deep Learning and Satellite Imagery to Quantify the Impact of the Built Environment on Neighborhood Crime Rates

Adyasha Maharana,[1] Quynh C. Nguyen,[2] Elaine O. Nsoesie[3]

**Affiliations:**
1. Department of Biomedical Informatics and Medical Education, University of Washington, Seattle, WA, United States
2. Department of Epidemiology and Biostatistics, University of Maryland, College Park, MD, United States
3. Institute for Health Metrics and Evaluation, University of Washington, Seattle, WA, United States

*Abstract: The built environment has been postulated to have an impact on neighborhood crime rates, however, measures of the built environment can be subjective and differ across studies leading to varying observations on its association with crime rates. Here, we illustrate an accurate and straightforward approach to quantify the impact of the built environment on neighborhood crime rates from high-resolution satellite imagery. Using geo-referenced crime reports and satellite images for three United States cities, we demonstrate how image features consistently identified using a convolutional neural network can explain up to 82% of the variation in neighborhood crime rates. Our results suggest the built environment is a strong predictor of crime rates, and this can lead to structural interventions shown to reduce crime incidence in urban settings.*

**Main Text:** Aspects of the built environment such as, building design, street layouts, land use, and environmental disrepair and desolation, have been associated with crime incidence. Differences in the density of these features have been linked to variations in crime rates across diverse geographical settings. However, different built environment characteristics influence particular types of crimes and may work through different mechanisms for crime inducement. For example, the presence of high schools, public parks, vacant lots or buildings can invite gang-related crimes (*1*, *2*). Also, urban features such as, public transit stations, convenience stores, bars and taverns, and high-rise buildings could increase neighborhood crime rates by attracting transient individuals, facilitating crime through alcohol and drugs, and concentrating disadvantage (*2–5*). Neighborhood crime has also been associated with structural aspects of the environment linked to the degree of accessibility, and the ease of entry and exit (*6*). Specifically, major transportation arteries and highways, intersections, alleys and mid-block connections that invite more traffic or enable easy escape have been associated with increased crime and risk of crime in both residential and commercial areas (*3*, *7–9*). Associations have also been noted between accessibility and higher crime when comparing blocks or street segments in high-crime and low-crime neighborhoods (*10*).

In contrast, structural changes in urban neighborhoods have been associated with a reduction in crime rates. For example, a study conducted in London observed that improving lighting in urban streets led to decreases in crime and increases in pedestrian street use after dark (*11*). In another community intervention in Sarasota, Florida, improvements in city lighting, landscaping, the addition of balconies or porches and residential units to commercial areas combined with new police initiatives for drug dealing and prostitution led to decreases in personal and property crime (*12*). Additional factors such as street configurations that reduce permeability of cars and cleaning and greening vacant properties were also associated with lower crime rates (*13*).

Although visually identifiable, quantifying the density of these environmental attributes across different geographic regions, populations and over time can be cumbersome. Studies linking neighborhood crime to features of the physical environment have heretofore been conducted using costly and time-consuming onsite visits to count relevant attributes (e.g., the number of liquor stores, vacant lots, and ratings of the level of graffiti or litter in the vicinity of interest) or neighborhood surveys to assess participant perceptions of their neighborhood. The resulting data can therefore be subjective since it relies upon participant or researcher perceptions, and assessment tools that vary across studies. Furthermore, sample sizes for most neighborhood studies tend to be small due to the burden of data collection. The absence of easily accessible data can hamper efforts to identify and quantify the impact of place on crime rates and other relevant public health measures.

Here, we demonstrate an accurate, scalable, and straightforward approach that combines a convolutional neural network model and satellite imagery to infer characteristics of the physical environment to assess the degree to which the physical environment can predict variations in crime rates ("predict" here does not indicate forecast of future events). We apply our method to predicting crime rates at the United States census tract level for three cities (Chicago, Illinois; St. Louis, Missouri; and Los Angeles, California) with high crime rates and available geo-referenced crime data. In contrast to existing methods, our approach is low cost, and can produce fine-grained estimates using publicly available data and software.

**Modeling Approach**

High-resolution satellite imagery are rich and comprehensive repositories of information for a variety of domains, ranging from crop health to the economy (*14–17*). Recent studies have shown that the application of deep neural networks to satellite images can enable characterization of the physical environment to study poverty, the economy and the demographic makeup of the United States (*18*, *19*). Deep neural networks and similar machine learning techniques can recognize landscape features from a bird's eye view and interpret them meaningfully. Deep neural networks can also provide crucial, scalable insights in a relatively inexpensive way.

However, in this rapidly evolving field, there have been no studies focused on the prediction of crime rates using data from satellite images. A related study adopted the Broken Windows theory (*20*) to identify city landscape features from Google Street View image for crime prediction but used the machine learning algorithm, support vector regression (*21*). This approach was tested for several US cities and achieved more than seventy percent accuracy in binary classification of areas with low and high rates of violent crime. Another study used multi-modal features to classify crime hot-spots in Chicago (*22*). In contrast, our approach is the first comprehensive assessment of the association between the built environment and overall numeric crime rates at the neighborhood level, achieved by applying deep convolutional neural networks to satellite images to extract predictive environmental features.

Our modeling approach involves three steps. First, we obtained geo-referenced time-stamped 2016 crime records provided by law enforcement departments for each of the cities (*23–25*). These data include both serious crimes and misdemeanors i.e. part I and part II offenses. The numbers of crime were aggregated to the census tracts in accordance with the boundaries established by the 2010 United States Census. As appropriate, some crimes were further separated into categories of *personal* (e.g., assault, battery, homicide) and *property* (e.g., robbery, property destruction) crime. We also obtained five-year estimates of socioeconomic and population characteristics from the American Community Survey (ACS). The number of crimes for each census tract was divided by the ACS population estimates to arrive at the number of crime incidents per 1,000 persons (hereafter referred to as "crime rates").

Next, we collected nearly 100,000 satellite images spanning each census tract from Google Static Maps API (application programming interface) at a zoom level of 18. These images were unlabeled. To overcome the challenge of working with unlabeled data, we used a transfer learning framework, similar to that used by Jean et al. (2016) for predicting poverty (*19*). We used a convolutional neural network which has been pre-trained on the ImageNet database to differentiate between 1,000 object categories. We fine-tuned the network to our specific problem of crime prediction by training the model on data comprised of images from census tracts with high and low crime rates defined as the top and lower fifteen percent of crimes rates. The updated model identified features pertinent to describing neighborhood structures (such as, green cover, buildings and roads), which could be useful for making meaningful associations between crime rates and the environment (see Fig. 1). A total of 4,096 features were extracted from the penultimate layer of the neural network. The convolutional neural network learns on its own without manual annotation of the data.

Finally, we fitted a regression model to evaluate how much the extracted features explain crime rates. To reduce the dimensionality of the feature matrix and identify relevant predictive features, we used elastic net, which is both a regularization and variable selection technique (*26*). Elastic net combines the advantages of Ridge regression and Least Absolute Shrinkage and Selection Operator (LASSO). We selected to use elastic net because we can identify relevant predictive features, while also keeping highly correlated predictors. We applied a rigorous model fitting approach involving a fivefold cross validation process, which involves splitting the data into five separate groups. Each group is used in the model fitting and the points (i.e., census tracts) not included at each iteration are used in prediction. For each of the three cities, we fitted individual models to predict crime rates in each city solely using the features extracted from satellite images. We also assessed how well our model predicts personal and property crimes. Next, we developed additional regression models for predicting crime rates using demographic and socioeconomic variables that have been studied extensively in the crime literature. We evaluated the models by comparing the root mean squared error (hereafter referred to as error) and the variation in crime rates explained by each of the models using the coefficient of determination ($r^2$).

**Results**

Our modeling approach achieved variable predictions of crime rates as measured at the census tract across the three US cities. Cross-validated predictions explained 36.16%, 59.96% and 81.59% of the variation in crime rates across census tracts in Los Angeles, Chicago and St. Louis, respectively. Overall, we observed that the high crime rates, which can be classified statistically as outliers, were mostly underestimated although typically predicted as the highest in the cities (see Fig. 2).

To investigate these deviations in predictions across the three cities, we developed separate regression models for low and median crime (hereafter referred to as the low crime model) for each city, and a single model for all high crime regions since the individual sample sizes for each city were small. The low crime models separately explained 77.42%, 59.43%, and 85.94% of the variation in crime rates across census tracts in Chicago (see Fig. 3), Los Angeles (see Fig. 4), and St. Louis (see Fig. 5). Furthermore, our high crime model explained 67.62% of the variation in the crime data, suggesting that census tracts with high crime rates have some similar predictive features. These predictive values were obtained despite not labeling the satellite images of locations with reported crimes.

To investigate whether our models were transferable between cities, we fitted a separate model using data for each city and then predicted crime rates for the other two cities. We found cross-predictions to be generally poor, which is most likely due to the unique distribution in crime rates across census tracts within each city, rather than differences in the predictive features. This suggests that separate models are needed to make reliable assessments of the association between features of the built environment and crime rates for each city.

To further assess how well our approach could predict different types of crime, we fitted separate models to personal and property crimes for each city. The model inclusive of all the personal crime data for each city explained 59.67%, 93.21% and 37.12% of variation in crime rates for the cities of Chicago, St. Louis and Los Angeles, respectively. With the exclusion of census tracts with high crime rates, the variance explained increased to 77.58% and 56.93% for

Chicago and Los Angeles respectively. Similarly, 46.65%, 90.23%, and 19.50% of the variation in property crime rates were explained for Chicago, St. Louis and Los Angeles by the model fitted to all census tracts. After excluding census tracts with high crime rates, the variation explained increased to 50.62% for Chicago and 23.22% for Los Angeles.

These observations suggest that our models are predictive of overall crime rates and do not favor personal or property crime. In addition, census tracts with high crime rates heavily influence predictions when models are fitted to all the data. Furthermore, we observed the highest accuracy in prediction for St. Louis, which has the highest deviation in crime rates across census tracts (standard deviation of 116.42, vs. 75.30 and 108.41 for Los Angeles and Chicago, respectively), probably making it easier for our modeling approach to differentiate between low and high crime census tracts.

To compare our findings to predictions based on socioeconomic and demographic variables, which have been extensively studied and associated with neighborhood crime rates, we developed regression models to predict crime rates based on variables related to unemployment, income, racial demographics, and education (*22*, *27–31*). Although we observed significant correlations between some of these variables and personal crime rates for Chicago, these associations were much weaker for the other cities. Specifically, variables highly associated with overall crime rates in our data included percent income below poverty ($\rho= 0.42$, where $\rho$ is the Pearson correlation), percent black population ($\rho=0.58$) and percent white population ($\rho= -0.54$). In addition, percent black population ($\rho=0.67$), percent income below poverty ($\rho=0.48$), percent white population ($\rho=-0.63$) and percent employed ($\rho=-0.52$) were also strongly associated with personal crime rates. These observations agree with reports on the distribution of crimes in Chicago (*32*). Similarly, the strongest positive predictors of overall crime rates in Chicago were poverty, percent black population, and percent population between the ages of ten and twenty. In contrast, the strongest negative predictors were population density, and employment. We observed similar associations for St. Louis and Los Angeles. Education was also negatively associated with crime rates in Los Angeles.

The model fitted to the entire socioeconomic and demographic dataset for each city explained 40.13%, 30.40% and 51.60% of the variation in crime rates across census tracts in Chicago, Los Angeles and St. Louis, respectively. After the removal of census tracts with high crime rates, the variation explained increased to 53.59% and 64.13% separately for Chicago and St. Louis. The model for census tracts with high crime rates across all cities, explained only 14.15% of the variance. These values are significantly lower than those observed for the model solely based on environmental features inferred from satellite imagery.

Furthermore, the socioeconomic factors explained 53.55%, 42.56% and 40.21% of the variance in personal crime rates for Chicago, Los Angeles and St. Louis, respectively. After excluding census tracts with high crime rates, the predictions improved slightly ($r^2 = 59.83\%$ and 49.48% respectively) for Chicago and Los Angeles, and significantly for St. Louis ($r^2 = 62.24\%$). The accuracy of predicting personal crime rates in Chicago using socioeconomic characteristics is comparable to that observed using environmental features suggesting that socioeconomic characteristics could be as strong a predictor of personal crimes as features of the physical environment. However, significant differences are noted in the predictions for St. Louis and Los Angeles.

Furthermore, the socioeconomic characteristics explained 25.80%, 54.20% and 25.15% of the variance in property crime rates for Chicago, St. Louis and Los Angeles, respectively. After excluding census tracts with high crime rates, the variance explained dropped slightly for Los Angeles ($r^2 = 22.01\%$), but increased for St. Louis ($r^2 = 63.36$) and Chicago ($r^2 = 41.99\%$). Overall, the socioeconomic characteristics do not strongly predict property crimes.

**Discussion**

Neighborhood crime rates can be explained by a complex interaction of environmental, societal, and individual level factors. While socioeconomic and demographic variables have been presented as predictors of crime, these factors do not completely explain neighborhood crime rates. In this study, we quantified the variation in crime rates at the census tract level across three cities that are explainable by features of the physical environment. Specifically, our approach demonstrates the use of deep neural networks to extract features of the built environment that are predictive of crime from high resolution satellite imagery. Notably, we demonstrate that detailed fine-grained estimates of neighborhood crime rates can be constructed with low cost data and tools.

Our results suggest that characteristics of the built environment are able to distinguish high and low crime areas above and beyond residential compositional characteristics. The differences in predictions across cities might indicate that for some cities the physical environment might explain variations in crime rates better than for others. Our results are in alignment with empirical research suggesting there is a relationship between physical disorder and fear of crime and crime rates (*33*, *34*). Thus, interventions to change the environment may help to prevent future crime, and improve individual-level health and psychological functioning. Crime and fear of crime have a myriad of potential health impacts including, worsening mental health, increases in substance use and drug overdoses (*33*, *35*) in addition to social and economic costs.

One potential approach used for environmental interventions is Crime Prevention Through Environmental Design (CPTED). CPTED is centered around incorporating design features that promote safety and security within a community (*15*). Design features include the following: natural surveillance, access control, territorial reinforcement, activity support, and maintenance. These design features help reduce opportunity for crime, increase social control, provide reassurance to community members by signaling to observers that disorder is not tolerated. Examples of specific community design strategies to help reduce crime and increase perceived safety include installing outside lighting to entrances, walkways, and parking lots;

decreasing visual barriers and concealed areas (e.g., underpasses); building fencing and walls to demarcate public and private property; designing landscaping with ground cover and tree canopy to allow for visibility and demonstrate ownership; increasing security systems; responding to maintenance issues (e.g., graffiti); and providing recreational facilities and structural support for safe activities.

However, there are some limitations to our modeling approach. First, we assume that our crime data is accurate. There is extensive criminology research suggesting that crime databases only represent a biased sample of all criminal offences (*36–40*). These data are influenced by several factors including, existing police priorities and crime incidence reporting. For example, although drug crimes tend to be widely distributed, police arrests on drug offences tend to be concentrated in lower income and high non-white population neighborhoods (*39*). In our data, we noted higher crime rates were reported in census tracts with lower income and with a higher percentage of blacks. Therefore, although our methods provide some quantitative association between the physical environment and crime rates, it should not be used as the sole predictor of crime rates. Also, processes are needed to address the bias inherent in these data.

The census tracts with some of the highest crime rates in Los Angeles and St. Louis include public parks. In spite of being low-populated areas, they register relatively high incidence of crime which may be attributed to the regular inflow of crowd from nearby areas. Crime rates for such areas can be better predicted by taking the structural features of surrounding census tracts into account. Some census tracts with very high crime rates in Chicago barely span across four to five blocks and also have a large margin of error in their population estimates. Prediction of crime for such areas also could benefit from wider coverage of neighborhood for feature extraction.

Work on deep learning methods continues and makes possible future improvements in feature extraction and predictions using these approaches. Our methods provide a state-of-the-art approach for automated extraction of features of the physical environment at a low cost to allow for the study of the association between the physical environment and crime rates, which can assist in the design and implementation of structural interventions shown to reduce crime incidence in urban settings.

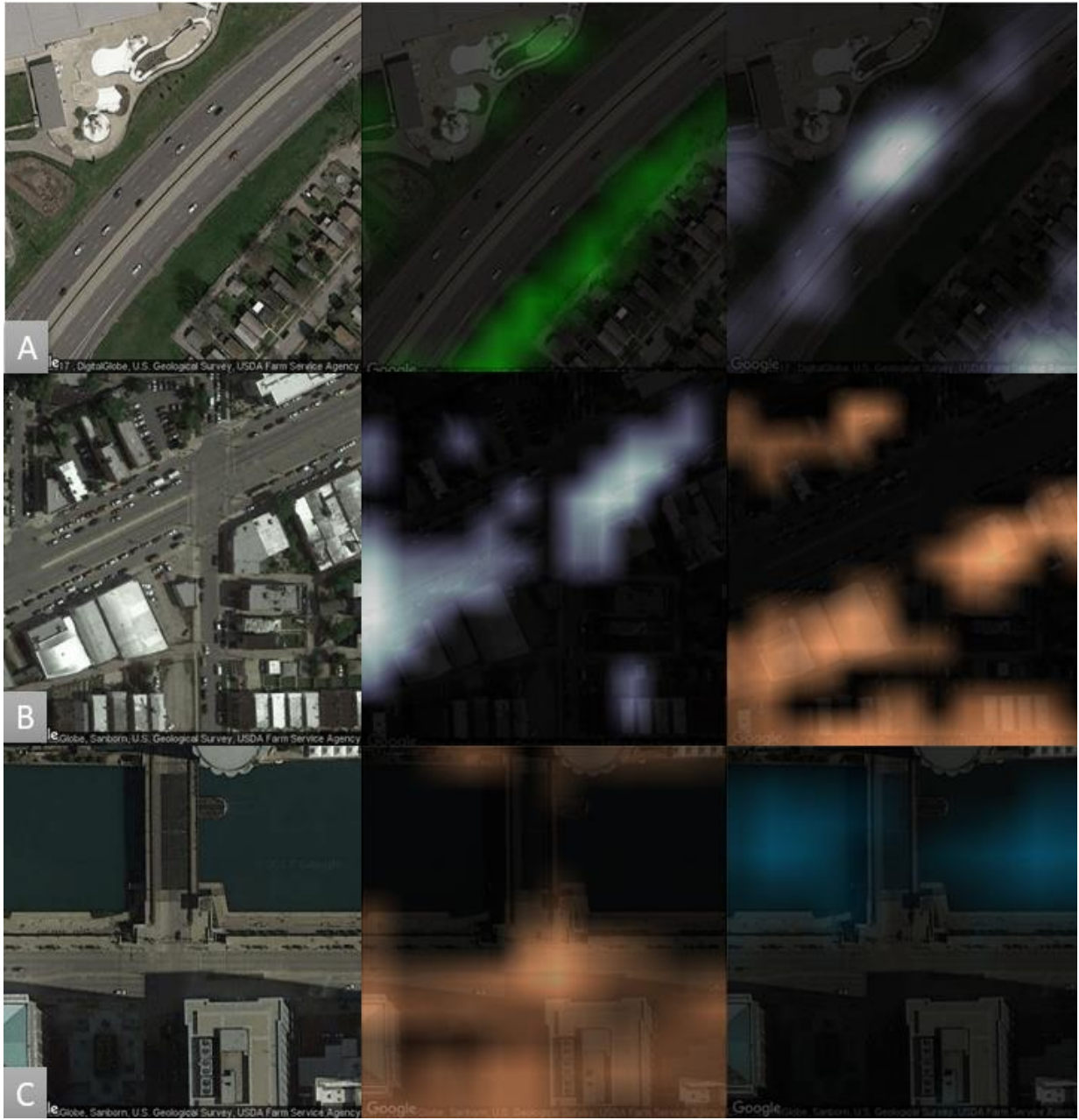

**Fig.1. Illustration of features identified by neural network model.** Filters from the convolutional network are interpolated to higher resolution and overlapped with corresponding images. The network learns to segment structural features from satellite images as shown in **(A), (B)** and **(C).** Extracted features represent lawns and highway in **(A),** roads and buildings in **(B),** water body and land in **(C).**

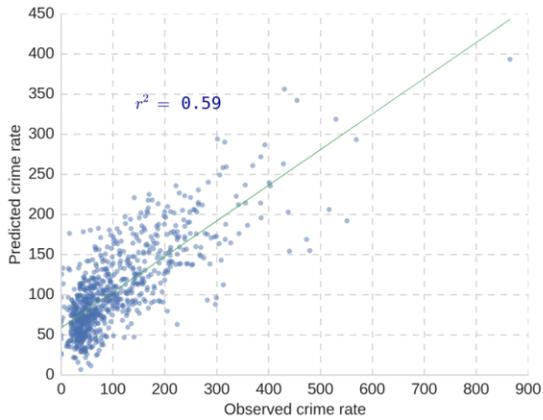
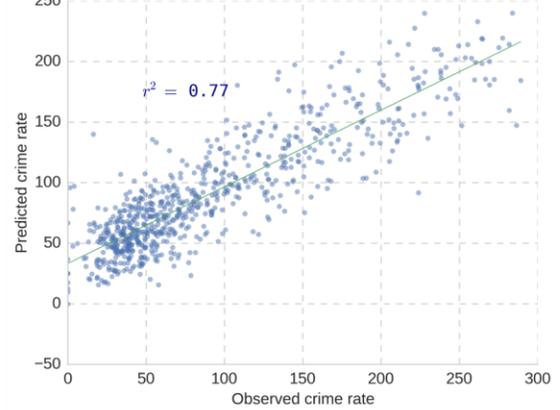
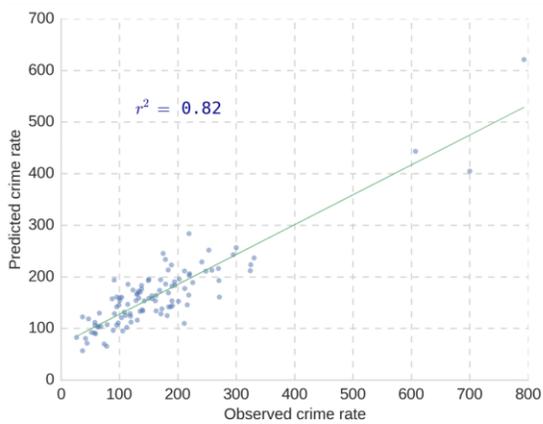
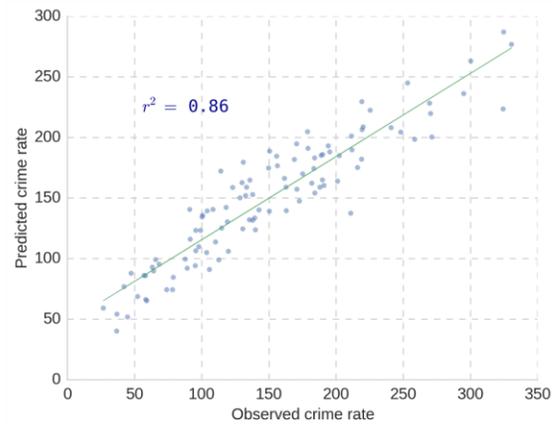
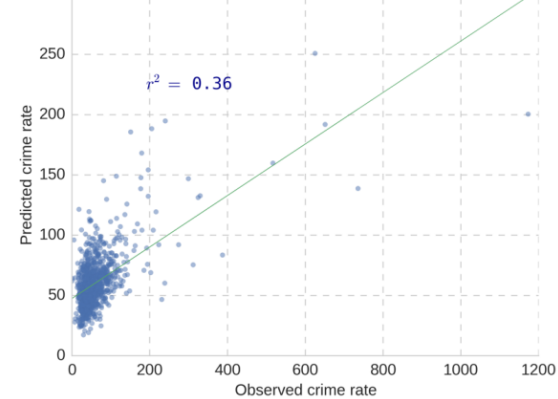
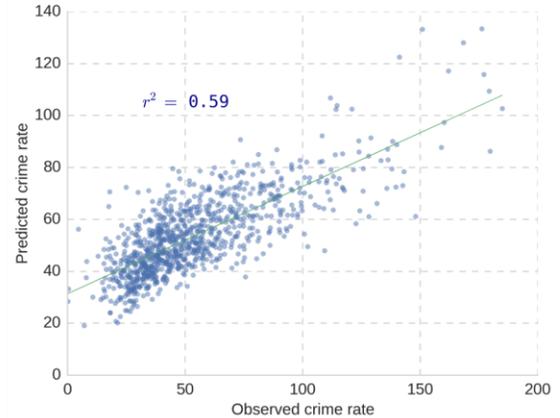

**Fig. 2. Predicted census-tract level crime rates (y axis) compared to the observed crime rates (x axis).** Results are shown for Chicago for the entire dataset (**A**) and the model without outliers (**B**). Similar results are presented for St. Louis, (**C**) and (**D**), and Los Angeles, (**E**) and (**F**). The $r^2$ values are based on fivefold cross-validation.

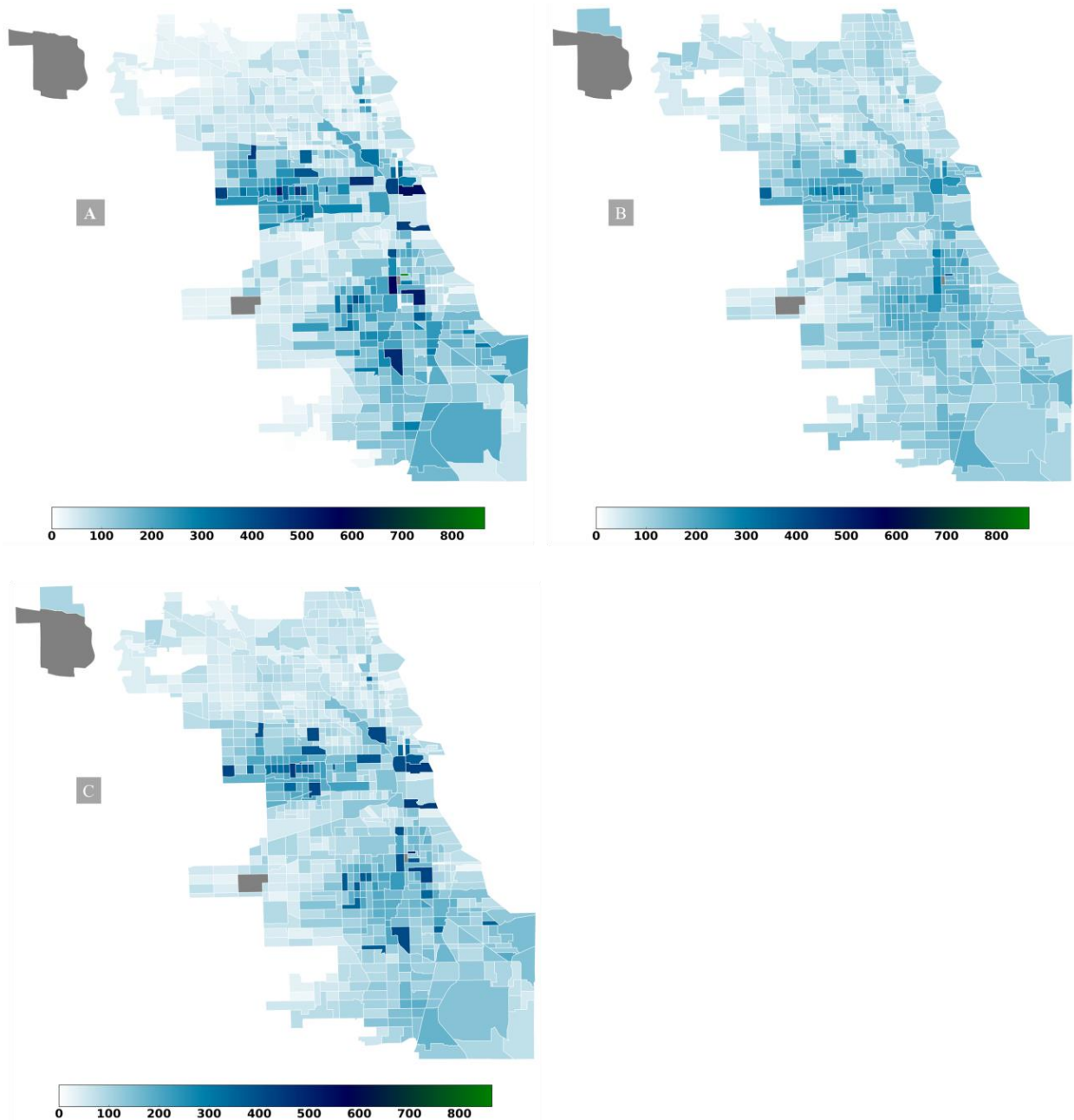

**Fig. 3. Predictions of crime rates across census tracts in Chicago, Illinois.** The figures show the actual crime rates per 1,000 persons **(A)**, predictions for a single model fitted to all crime levels **(B)** and the combined predictions from the model for low and median crime region, and model for high crime rates **(C)**. The gray shaded regions are census tracts with zero population or no reported crimes.

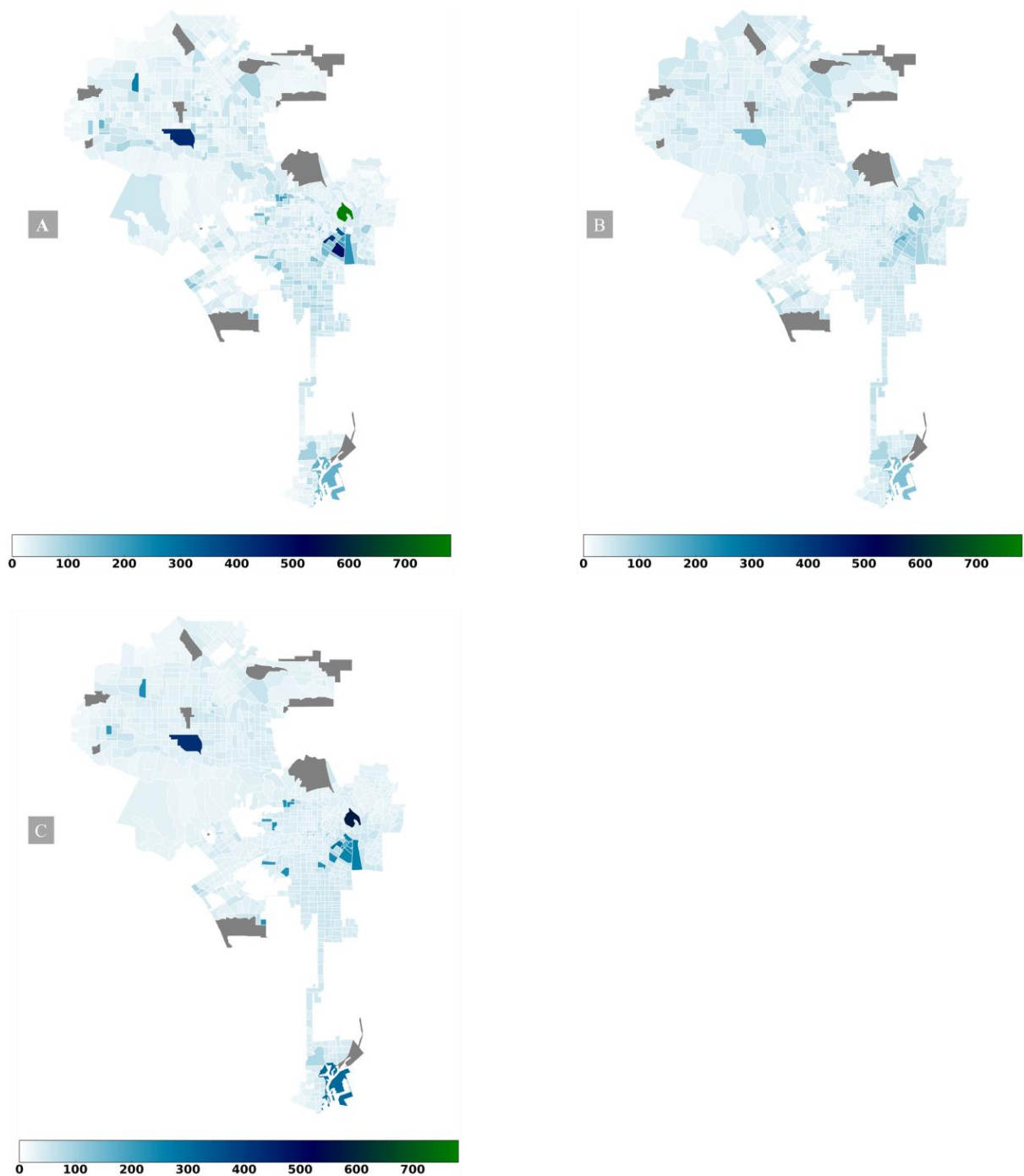

**Fig. 4. Predictions of crime rates across census tracts in Los Angeles, California.** The figures show the actual crime rates per 1,000 persons **(A)**, predictions for a single model fitted to all crime levels **(B)** and the combined predictions from the model for low and median crime region, and model for high crime rates **(C)**. The predictions improve when outliers (i.e., census tracts with the highest crime rates) are excluded from the model.

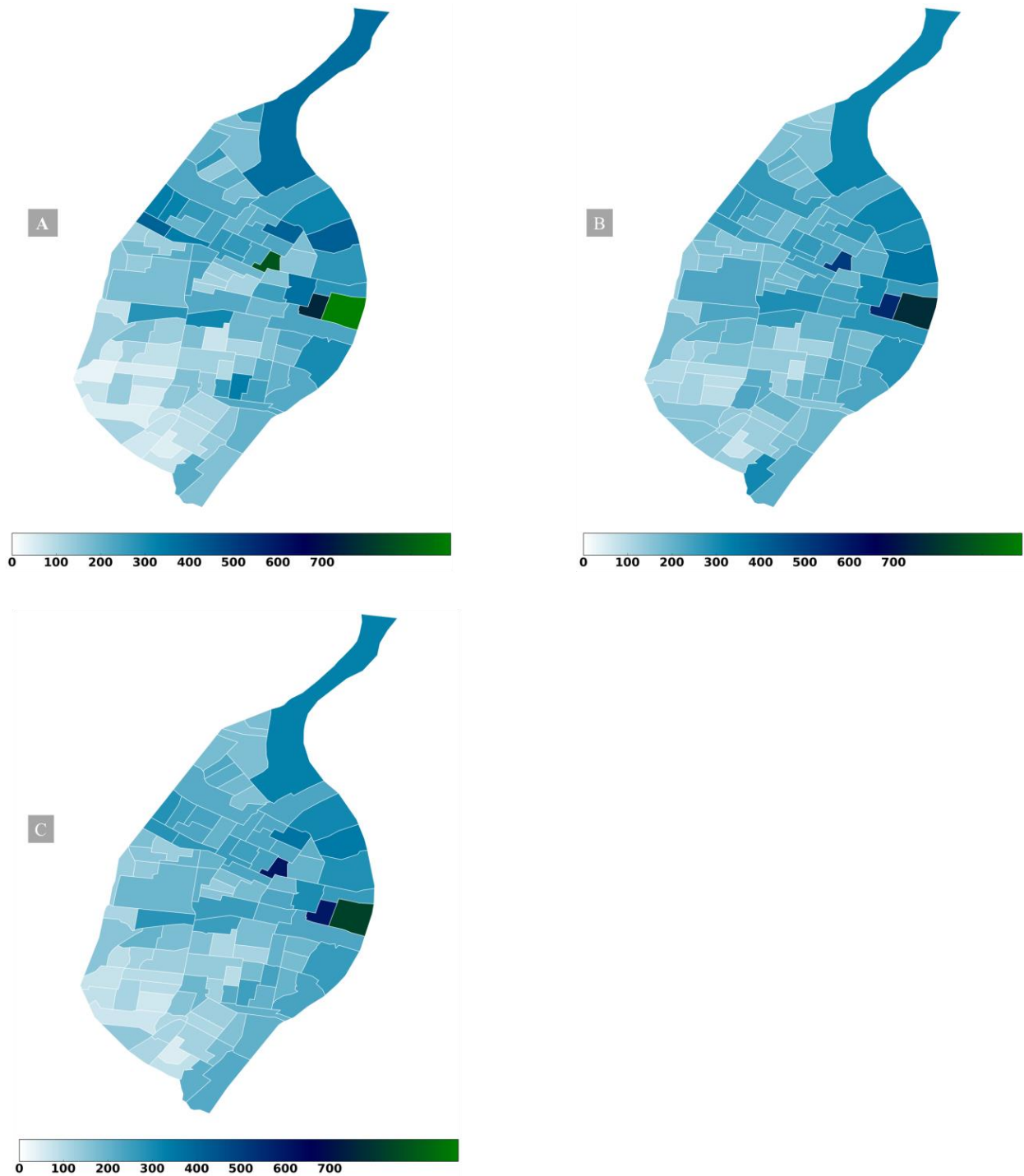

**Fig. 5. Predictions of crime rates across census tracts in St. Louis, Missouri.** The figures show the actual crime rates per 1,000 persons **(A),** predictions for a single model fitted to all crime levels **(B)** and the combined predictions from the model for low and median crime region and model for high crime rates**(C).** Predictions in (B) and (C) are almost identical.